\documentclass[aps,prd,twocolumn,amssymb,showpacs,superscriptaddress]{revtex4}
\usepackage{graphicx,bm,color}
\usepackage{amsmath}
\usepackage{amssymb}
\usepackage{amsfonts}

\newcommand{\be}{\begin{equation}}
\newcommand{\ee}{\end{equation}}
\newcommand{\bea}{\begin{eqnarray}}
\newcommand{\eea}{\end{eqnarray}}
\newcommand{\beaa}{\begin{eqnarray*}}
\newcommand{\eeaa}{\end{eqnarray*}}

\newcommand{\nn}{\nonumber \\}
\newcommand{\e}{\mathrm{e}}

\begin{document}

\title{Spontaneous symmetry breaking in cosmos: The hybrid symmetron \\
as a dark energy switching device}

\author{K.~Bamba}
\email{bamba@kmi.nagoya-u.ac.jp}
\affiliation{Kobayashi-Maskawa
Institute for the Origin of Particles and the Universe, \\
Nagoya University, Nagoya 464-8602, Japan}
\affiliation{Physics Division,
National Center for Theoretical Sciences, Hsinchu, Taiwan 300}

\author{R.~Gannouji}
\email{gannouji@rs.kagu.tus.ac.jp}
\affiliation{Department of Physics, Faculty of Science,
Tokyo University of Science,
Tokyo 162-8601, Japan}

\author{M.~Kamijo}
\email{kamijo@th.phys.nagoya-u.ac.jp}
\affiliation{Department of Physics, Nagoya University, Nagoya
464-8602, Japan}

\author{S.~Nojiri}
\email{nojiri@phys.nagoya-u.ac.jp}
\affiliation{Kobayashi-Maskawa
Institute for the Origin of Particles and the Universe, \\
Nagoya University, Nagoya 464-8602, Japan}
\affiliation{Department of
Physics, Nagoya University, Nagoya 464-8602, Japan}

\author{M.~Sami}
\affiliation{Department of Physics, Nagoya University, Nagoya
464-8602, Japan}
\affiliation{Centre for Theoretical Physics,
Jamia Millia Islamia, New Delhi-110025, India}

\begin{abstract}

We consider symmetron model in a generalized background with a hope
to make it compatible with dark energy. We observe a ``no go'' theorem at
least in case of a conformal coupling. Being convinced of symmetron
incapability to be dark energy, we try to retain its role for
spontaneous symmetry breaking and assign the role of dark energy
either to standard quintessence or $F(R)$ theory which are switched
on by symmetron field in the symmetry broken phase. The scenario
reduces to standard Einstein gravity in the high density region.
After the phase transition generated by symmetron field, either the
$F(R)$ gravity or the standard quintessence are induced in the low
density region. we demonstrate that local gravity constraints and
other requirements are satisfied although the model could generate
the late-time acceleration of Universe.

\end{abstract}


\maketitle

\section{Introduction \label{S1}}

One of the inconsistencies of hot big bang model is associated with
the observed primordial density fluctuation necessary for large
scale structure in Universe. The paradigm which successfully
addresses this problem along with the resolution of other
theoretical issues such as flatness and horizon problems is
{\it inflation}. The standard model of Universe is also plagued with age
crisis which is late time inconsistency. In the standard model, the
only known way to address this problem is provided by late time
acceleration of Universe. This serves as an indirect evidence of the
presence of an exotic matter of repulsive nature in Universe dubbed
{\it dark energy} in the framework of standard lore. The discovery
of the accelerating expansion of Universe \cite{Riess,Perlmutter} in
1998 supports this hypothesis. These observations tell us that the
dark energy amounts to about 73\% of the total energy budget of the
Universe at present
\cite{Kowalski,vpaddy,review2,review3,review3C,review3d,review4,review5,Bamba:2012cp}.

The simplest cosmological model of dark energy is based upon
cosmological constant ($\Lambda$) but the framework faces the most
difficult conceptual and theoretical issues associated with
($\Lambda$). With a hope to alleviate some of these problems, a
variety of scalar field models such as quintessence, phantoms,
k-essence and tachyons have been investigated in the literature.
\cite{Peebles:1987ek,Ratra:1987rm,Chiba:1997ej,Zlatev:1998tr,Peebles:1998qn}.
Naively, the mass of slowly rolling quintessence field should be of
the order Hubble rate in the present Universe, $H_0 \sim
10^{-33}\,\mathrm{eV}$. In case, it is much larger (smaller) than
$H_0$, it would mimic stiff matter ($\Lambda$). The incredible small
mass scale of dark energy gives rise to formidable problems. In
case, the scalar field is coupled to matter, as it should be in
general, the propagation of the scalar field could generate the
large correction to the Newton law and could be excluded by the
local physics constraints.

In order to address the problem related to local gravity
constraints, three mechanisms of mass screening have been employed
in cosmology. (1) Chameleon scenario: This mechanism operates with
the field mass which becomes dependent on the local density of
environment such that the latter gets large in high density regime
thereby leading to effectively decoupling of field from matter or
suppression of fifth force \cite{Khoury:2003aq,Khoury:2003rn}. (2)
Vainshtein mass screening \cite{claudia}: This mechanism is superior
to chameleon and operates dynamically with non-linear derivative
interactions. In the neighborhood of a massive body, the non-linear
kinetic terms become strong leading to the decoupling of the field
from the source in a large region, around the massive body,
specified by the so called Vainshtein radius.

Recently, a very interesting, third screening mechanism closely
related to chameleon dubbed {\it symmetron} is proposed in
Refs.~\cite{Olive:2007aj,Hinterbichler:2010es,Hinterbichler:2011ca}.
The idea of symmetron is related to the late time cosmic phase
transition via spontaneous symmetry breaking. Similar to the
chameleon model, the mass of the symmetron scalar field depends on
the density of environment in a specific way due to its direct
coupling to matter. For densities lower than some critical value,
symmetron becomes tachyon and the symmetric vacuum state ($\phi=0$)
is no longer a true vacuum. In this case, there are two vacua with
$Z_2$ symmetry which breaks soon after we choose one of these. The
mass around the true minimum is well behaved. It is clear that if
symmetron is to be relevant to the dark energy, the phase transition
should take place when the density of environment is low and the
mass of the symmetron in the true vacuum is around $H_0$, which
could be the first requirement. The second requirement is that the
symmetron should be invisible locally, that is, the fifth force
induced by the symmetron should be negligibly small as compared to
the Newtonian force of gravity. The second requirement can be easily
satisfied in the original symmetron model. As for the first,
however, local physics imposes severe constraints on the symmetron
mass which turns out to be quite heavy to derive late time
acceleration. Although the symmetron might play some role during
structure formation, the original idea of introduction of symmetron
seems to be defeated.


No doubt that the symmetron presents a beautiful idea and we believe
that beauty cannot go for waste. Though the original idea of the
symmetron does not seem to work for dark energy without unnatural
fine tunings but the role of symmetron in cosmic symmetry breaking
could be retained. We propose a model where we use the symmetron to
felicitate the cosmic phase transition and the role of dark energy
is played either by a second quintessence field or $F(R)$ gravity
which are switched on by the symmetron after the phase transition is
over. In our proposal, the action reduces to that of the $F(R)$
gravity \cite{Capozziello:2002rd,Carroll:2003wy,Nojiri:2003ft} (for
reviews, see \cite{Nojiri:2006ri,Nojiri:2010wj,Bamba:2012cp}) or standard
quintessence field $\varphi$ in the bulk, where the energy density
is very small but action reduces to the usual Einstein-Hilbert one
in the high density region, like in/on the earth, in the solar
system, and in the galaxies, which makes the model consistent with
the observations and local experiments.

The plan of the paper is as follows. We, first, very briefly
revisit the standard symmetron scenario, specially focusing on its
failure for dark energy. As an attempt to to avoid the problem, we
investigate a model where symmetron couples to curvature scalar but
observe a no go theorem for late
time acceleration. Hence we propose a model where
$F(R)$ gravity couples with symmetron and show that there exists a
parameter region which satisfies possible constraints.
The last section is devoted to summary and discussion.

\section{Standard Symmetron \label{S2}}

In this section, we briefly review the standard symmetron model.
This model is a generic attempt to implement the original
Ginsburg-Landau idea of phase transitions in cosmology. The model
is based upon Higgs type scalar field potential with $Z_2$
symmetry. In the high density regime, the system resides in the
symmetric ground state, symmetry breaks spontaneously in low density
region. The consistency of the model with local physics severely
constraints the symmetron field to be relevant to dark
energy \cite{Hinterbichler:2010es,Hinterbichler:2011ca}.

\subsection{Failure of symmetron to be dark energy \label{S3}}

Let us very briefly outline the basic features of the original
symmetron scenario based upon the following Einstein
frame \cite{Olive:2007aj,Hinterbichler:2010es,Hinterbichler:2011ca}
\begin{align}
\label{eq:action}
\mathcal{S}=\int d^4x\sqrt{-g}\left[
\frac{M_\mathrm{Pl}^2}{2}R-\frac{1}{2}(\nabla\phi)^2-V(\phi) \right]\nonumber\\
+\mathcal{S}_m\left[ A^2(\phi)g_{\mu\nu},\Psi_m \right]\, .
\end{align}
Here
$\mathcal{S}_m$ expresses the action of matter denoted by $\Psi_m$.
We also choose $A(\phi)=1+\left[\phi^2/\left(2M^2\right)\right]$,
where $M$ is a mass scale in the model. The effective potential then
takes the following form
$
V_\mathrm{eff}=\left(1/2\right)\left(\rho/M^2-\mu^2\right)\phi^2
+\left(1/4\right)\lambda\phi^4
$.
The mass of the field now depends upon the density of environment,
naively, the field mass is given by,
$m_\mathrm{eff}^2={\rho}/{M^2}-\mu^2$. In high density regime, the
mass depends upon density linearly, $m_\mathrm{eff}^2 \sim
\rho/M^2>0$. In this case, the system resides in the symmetric
vacuum $\phi=0$. The requirement of local gravity constraints puts
an upper bound on $M$ and there is no a priori reason for it to be
consistent with dark energy. In case of chameleon, there is
more flexibility, the mass depends on density non-linearly.
As shown in \cite{Hinterbichler:2010es,Hinterbichler:2011ca}, $M \leq
10^{-4}M_\mathrm{Pl}$ in case of symmetron. As the density redshifts
with expansion and $\rho$ falls below $\mu^2M^2$, tachyonic
instability builds in the system and the symmetric state $\phi_0=0$
is no longer a true minimum; the true minima are then given by
$
\phi_0=\pm \sqrt{\left[\mu^2-\left(\rho/M^2\right)\right]/\lambda}
$,
and the mass of the symmetron about the true minimum is,
$m_\mathrm{s}=\sqrt{2}\mu$ (at low density). Universe goes through a crucial
transition when late time acceleration sets in around the redshift
$z\sim 1$. One thus assumes that the phase transition or symmetry
breaking takes place when $\rho$ is around $\rho_\mathrm{cr}$ as
$
\rho_\mathrm{cr} \simeq M^2\mu^2 \to \mu^2 \simeq
\left(H_0^2 M_\mathrm{Pl}^2\right)/M^2 \to m_\mathrm{s}\simeq
\left(H_0 M_\mathrm{Pl}\right)/M
$.
This means that $m_\mathrm{s}\ge 10^4H_0$ which is larger than the
required quintessence mass by four orders of magnitude. In this
case, the field rolls too fast around the present epoch making
itself untenable for cosmic acceleration. Invoking the more
complicated potential with minimum with the required potential height
does not solve the problem; field goes oscillating for a long time
and does not settle in the minimum unless one arranges symmetry
breaking very near to $z=0$ by unnatural fine tuning of
parameters \cite{Hinterbichler:2010es,Hinterbichler:2011ca}. This
also undermines the beauty of the underlying theory which is
renormalizable and has an edge over chameleon theories which use
potentials with complicated functional forms. In what follows we
shall present a model which uses the $\lambda \phi^4$ theory in a
modified background.

\subsection{An effort to make symmetron compatible
with dark energy that does not succeed \label{S4}}

Since the complicated choice of original potential will not help. We
thus modify the effective potential by virtue of the modification of
the background. We consider the following model,
\begin{align}
\label{eq:action2}
\mathcal{S}=\int
d^4x\sqrt{-g}\left[F(\phi)R-\frac{1}{2}(\nabla\phi)^2-V(\phi)\right]\nonumber\\
+\mathcal{S}_m\left[A^2(\phi)g_{\mu\nu},\Psi_m\right]\, ,
\end{align}
where $F(\phi)$ is an arbitrary function of the field which should
satisfy the obvious requirements of $Z_2$ symmetry and the
requirement that the model should not
have instabilities. It is interesting to notice that if $F=A^2$,
the model (\ref{eq:action2}) is equivalent to
a minimally coupled quintessence field.

Variation of action with respect to $g_{\mu\nu}$ and $\phi$ gives the
following equations of motion
\begin{align}
\label{mEinsteineq1}
F(\phi) G_{\mu\nu} = &\frac{1}{2}\partial_\mu \phi
\partial_\nu \phi-\frac{1}{2}g_{\mu\nu}\Big[\frac{1}{2}
(\nabla \phi)^2+V(\phi)\Big]\nonumber\\
&+\nabla_{\mu\nu}F-g_{\mu\nu}\Box F
+\frac{1}{2}T_{\mu\nu} \, ,\\
\Box\phi = & V'(\phi)-R F'-\frac{A'}{A}T \, .
\label{mphieq1}
\end{align}
Taking the trace of Eq.~(\ref{mEinsteineq1}),
we can express $R$ through $T$ and combinations of $\phi$ and
substituting the same in the field equation gives
\begin{align}
\label{mphieq2}
\left(F+3F'^2\right)\Box \phi &
+ F'\left[3F''+\frac{1}{2}\right] (\nabla \phi)^2\nonumber\\
&= \left(\frac{F'}{2} -\frac{A'}{A}F\right)T+FV'-2F'V\, .
\end{align}
The second term in Eq.~(\ref{mphieq2}) is an extra kinetic term
obtained due to field coupling with curvature. We should ensure
that it does not give rise to instabilities.
For simplicity, it is better to switch it off
but keeping modification of gravity alive,
\begin{equation}
\label{FF1}
F'\left(3F''+\frac{1}{2}\right)=0\, .
\end{equation}
In (\ref{FF1}), an option $F'=0$ corresponds to the general
relativity by Einstein, which is not desirable for us.
The second option is interesting,
\begin{equation}
\label{FF2}
3F''+\frac{1}{2} =0 \Rightarrow F(\phi)
= \alpha+\beta\phi-\frac{\phi^2}{12}\, .
\end{equation}
Since we like to have $Z_2$ symmetry to hold, we choose
$\beta=0$ and for $\alpha$, the obvious choice is,
$\alpha=M_\mathrm{Pl}^2/2$.
We then have the expression for $F(\phi)$,
\begin{equation}
\label{FF3}
F(\phi)=\frac{M_\mathrm{Pl}^2}{2}-\frac{\phi^2}{12}\, .
\end{equation}
We notice that $F(\phi)>0$ in generic range of $\phi$ and corresponds
to the conformal coupling.
The field equation with this choice becomes
\begin{equation}
\label{mphieq3}
\frac {M_\mathrm{Pl}^2}{2}\Box \phi= \left(\frac{F'}{2}
 -\frac{A'}{A}F\right)T+FV'-2F'V\, .
\end{equation}
We now make use of generic functional forms of $V(\phi)$ and
$A(\phi)$,
\begin{eqnarray}
\label{FF4}
V(\phi)=-\frac{1}{2}\mu^2\phi^2+\frac{1}{4}\lambda \phi^4\, ,\quad
A(\phi)=1+\epsilon\frac{\phi^2}{2M^2}\, .
\end{eqnarray}
The field equation (\ref{mphieq3}) then takes the following form
$\Box \phi=dV_\mathrm{eff}/d\phi$ with
\begin{align}
\label{veff}
V_\mathrm{eff}&=\left[-\frac{1}{2}\mu^2+ \frac{\rho}{2}
\left(\frac{\epsilon}{M^2}+\frac{1}{6M_\mathrm{Pl}^2}\right) \right]\phi^2 \nonumber\\
&\qquad +\frac{1}{4}\left(\lambda-\frac{\mu^2}{6 M_\mathrm{Pl}^2}
 -\frac{\epsilon \rho}{12M^2 M_\mathrm{Pl}^2}\right)\phi^4\, .
\end{align}
In order to carry out the consistency check, it is instructive
to look at the expression of $V_\mathrm{eff}$ in low and high density
regimes,
\begin{widetext}
\begin{align}
\label{FF5}
& \mbox{High density regime}: V_\mathrm{eff} \simeq \frac{\rho}{2}
\left[\left(\frac{\epsilon}{M^2} +\frac{1}{6M_\mathrm{Pl}^2}\right)
\phi^2-\frac{\epsilon}{24M^2M_\mathrm{Pl}^2}\phi^4\right] \, , \\
\label{FF6}
& \mbox{Low density regime}: ~ V_\mathrm{eff} \simeq -\frac{1}{2}\mu^2\phi^2
+ \frac{1}{4}\left(\lambda-\frac{\mu^2}{6M_\mathrm{Pl}^2}\right)\phi^4\, .
\end{align}
\end{widetext}
Stability then demands that the effective potential should be
bounded from below in any regime or the coefficient of $\phi^4$
should be positive. Secondly the coefficient of the kinetic energy
term should also be positive necessary to avoid the ghost
instabilities. And these requirements translate into restrictions on
parameters, namely, $\epsilon <0$ (without no loss of generality we
can take $\epsilon=-1$), $1-\frac{\mu^2}{6 M_\mathrm{Pl}^2}>0$,
$M\ge \sqrt{6}M_\mathrm{Pl}$.

Let us now consider spontaneous symmetry breaking which takes place
when the coefficient of $\phi^2$ in Eq.~(\ref{veff}) turns negative
which happens when
\begin{equation}
\label{FF7}
\rho_\mathrm{SB}={\mu^2}\left({\frac{1}{6M_\mathrm{Pl}^2}
 -\frac{1}{M^2}}\right)^{-1}\, .
\end{equation}
Since we like to have the phase transition to take place at late
times when Universe enters the regime of accelerated expansion,
we shall identify $\rho_\mathrm{SB}$ with $\rho_\mathrm{cr}$
\begin{equation}
\label{FF8}
\rho_\mathrm{SB}\simeq \rho_\mathrm{cr}=H_0^2M_\mathrm{Pl}^2 \to \mu^2\simeq
\left(\frac{1}{6M_\mathrm{Pl}^2} -\frac{1}{M^2}\right)H_0^2M_\mathrm{Pl}^2\, .
\end{equation}
Since the effective mass of the field in the low density
regime is $m_\mathrm{s}=\sqrt{2}\mu$ and we like to have it to be of
the order of $ H_0$, we define
$\mu=\alpha H_0$  where $\alpha=O(1)$.
Using the relation, $M>\sqrt{6}M_\mathrm{Pl}$, we notice that
\begin{equation}
\label{FF9}
\alpha^2H_0^2\simeq
\left(\frac{1}{6M_\mathrm{Pl}^2}-\frac{1}{M^2}\right)H_0^2M_\mathrm{Pl}^2
\to 0<\alpha<1/\sqrt{6}\, .
\end{equation}
It is important to check whether or not the model is consistent with
local physics or the local gravity constraint on symmetron mass are
compatible with dark energy. In order to do that, we need to
transform the action (\ref{eq:action2}) into the Einstein frame and
obtain expressions for transformed potential and the coupling. It is
then not difficult to demonstrate that the  local gravity
constraints in this case impose $\alpha>10^4$ which do not match the
cosmological condition derived previously (\ref{FF9})

Let us try to understand why it happens. Broadly, the modified
background brings two changes from the point of view of Einstein
frame. First, the effective potential, secondly, the coupling $A^2$
in the matter Lagrangian. Both terms are drastically modified but in
the leading approximation, $\phi/M \ll 1$ $\left(
\phi/M_\mathrm{Pl}\ll 1 \right)$, we recover the standard symmetron
action. Thus roughly, the mass scale $M$ got replaced by
$\left(-1/M^2+1/6M^2_\mathrm{Pl}\right)^{-1}$ which also defines the
mass of symmetron in the broken phase, see Eq.~(\ref{FF7}), thereby
leading to the same local gravity constraint as in the original
symmetron model. Hence we see that the only difference would be a
renormalization of the coefficients in the model, but the
constraints will remain same, and therefore the symmetron cannot
play the role of a dark energy fluid.  It is important to say few
words about this generic feature of modified gravity models.


Recently, a no go theorem related to the scope of modified theories
based upon chameleon/symmetron was discussed in
Ref.~\cite{Wang:2012kj}. We have not tried to avoid the ``no go''
theorem, our proposal is rather in agreement with
Ref.~\cite{Wang:2012kj} that late time acceleration is not the
result of gravity modification but is driven by some quintessence
field or a cosmological constant. In the subsection to follow, we
add some clarifications on the important findings of
Ref.~\cite{Wang:2012kj} in the background of our proposal.

\section{Scope of chameleon/symmetron supported modification of gravity
for self acceleration \label{S15}}

There are two ways of gravity modification: (1) Apart from the
massless spin-2 object, there is essentially at least an extra
scalar degree of freedom which is exchanged apart from the graviton.
In order to be relevant at large scales where modification is
sought, the latter should gives rise to the effect of same strength
as that gravitational interaction. This would then cause a havoc
locally. The screening mechanism mentioned above address this
problem by locally suppressing the exchange effects of the scalar
degree of freedom. (2) The conformal coupling $A(\phi)$ also modifies
the strength of gravitational interaction. To pass the local tests,
$A(\phi)$ should be very closely equal to one in high density regime
in theories based upon chameleon/symmetron mechanism. As mentioned
above, Universe has undergone a phase transition during $0<z<1$
which is a large scale phenomenon and one might think that the
screening which is a local phenomenon should not impose severe
constraints on how $A(\phi)$ changes during the period acceleration
sets in. It, however, turns out that the change the conformal
coupling suffers as redshift changes from one to zero is negligibly
small. Then a question arises, can such a conformal coupling be
relevant to late time acceleration? It is well known that the de
Sitter Universe is conformally equivalent to the Minkowski
space-time. Has the conformal transformation changed physics? By
`physics', we mean the relationship between physical observables
which is same in both the frames. In the Einstein frame we have the
Minkowski space-time in which there is a scalar field sourced by the
conformal coupling which is dynamical and couples to matter
directly. The masses of all material particles are time dependent by
virtue of $A(\phi)$. As a result, one would see the same relations
between physical observables in both the frames. The acceleration
dubbed {\it self acceleration} is generic which can be removed
(caused) by virtue of conformal coupling \cite{Wang:2012kj}. Similar
thing happens with Dvali-Gabadadze-Porrati (DGP) model
\cite{Dvali:2000hr,Deffayet:2001pu}. Late time cosmic acceleration
which cannot be affected by conformal coupling is caused by the
presence of slowly rolling (coupled) quintessence and is not a
generic effect of modified theory of gravity.

Let us briefly check how it happens.

We have the following translation between the Einstein and Jordan
frames,
\begin{equation}
\label{EJ1}
a^J(t^J)=A(\phi)a^E(t^E)\, ,\quad dt^J=A(\phi)dt^E\, ,
\end{equation}
and conformal time which is same in both the frames, $dt=a(t)d\eta$.
In (\ref{EJ1}), $a^J$ $(a^E)$ is a scale factor and $t^J$ $(t^E)$ is the cosmic time
in the Jordan (Einstein) frame, respectively.

Following Ref.~\cite{Wang:2012kj}, it is easy to check that,
\begin{equation}
\label{EJ5}
\ddot{a}^Ja^J-\ddot{a}^Ea^E
=\left(\frac{A''}{A}-\frac{A'^2}{A^2}\right)
=\left(\frac{A'}{A}\right)'\, .
\end{equation}
On the right hand side. ``prime'' $(\ \ '\ )$ denotes the derivative with
conformal time in the Einstein frame. Though the conformal time is same in
both frames but coupling is defined only in the Einstein frame.

Let notice that acceleration in the Einstein frame cannot be caused by
conformal coupling,
\begin{equation}
\label{EJ6}
\frac{\ddot{a}}{a}=-\frac{1}{6M_\mathrm{Pl}^2}
\left((\rho_\phi+3P_\phi)+\beta_s\rho A(\phi)\right)\, .
\end{equation}
It is clear that in case acceleration takes place in the Einstein frame,
it can only be caused by slowly rolling quintessence ($\rho_\phi+3P_\phi<0$).
This implies that acceleration in the Jordan frame and no acceleration
in the Einstein frame is generic effect of conformal coupling or gravity
modification. In this case,
while passing from the Jordan to the Einstein frame, the acceleration is
completely removed, its affects in the Einstein frame are contained in
the coupling such that the relationship between physical observables
are same in both the frames.
The definition of self acceleration \cite{Wang:2012kj},
$\ddot{a}^Ea^E<0$ ($\ddot{a}^Ja^J>0)$, then implies
\begin{equation}
\label{EJ7}
\left(\frac{A'}{A}\right)\ge \ddot{a}^Ja^J\, .
\end{equation}
Since, $A'=\dot{a}^J \Delta A$
($\Delta A$ -- change over one Hubble
(Jordan) time), it follows that
\begin{equation}
\label{EJ8}
a^J \frac{d}{dt^J}\left(\dot{a}^J \frac{\Delta A}{A}\right)\ge
a^j\ddot{a}^J\, ,\quad
\Delta A=\left(\frac{1}{H^J}\frac{dA}{dt^J}\right)\, .
\end{equation}
Integrating left right the above relation, we get \cite{Wang:2012kj}
\begin{equation}
\label{EJ9} \frac{\Delta{A}}{A}\gtrsim 1\, .
\end{equation}
This quantity gives formally change of $A$ over one Hubble time in
the Jordan frame. As mentioned earlier, $A$ is defined in the Einstein frame
and naturally, the estimates of $A$ from screening are obtained in
that frame
\begin{equation}
\label{EJ10}
\left(\frac{\Delta A}{A}\right)_J
=\frac{\left(\frac{1}{H^E}\frac{ dA}{dt^E} \right)/A}
{ 1+\left(\frac{1}{H^E}\frac{ dA}{dt^E} \right)/A}\, .
\end{equation}
In case, $\Delta A$ is small or is of the order of one in the Einstein
frame, it will be so in the Jordan frame.

As demonstrated in Ref.~\cite{Wang:2012kj},
screening imposes a severe constraint on
the change of coupling during the last Hubble time, $\Delta A \ll 1$.
Thus self acceleration cannot take place in this case. In most of
the models supported by chameleon/symmetron screening, acceleration
takes place in both frames such that $\ddot{a}^Ja^J$ and
$\ddot{a}^Ea^E$ cancel each other with good accuracy or
$\Delta A \ll 1$. In this case acceleration can only be caused by slowly rolling
quintessence. In $F(R)$ theories, the scalar field and the coupling
both are made of $F'(R)$ which imbibes the gravity modification.
Since screening does not allow the conformal coupling to felicitate
self acceleration, the problem simply reduces to coupled
quintessence which one could deal with without really invoking
$F(R)$ theories. It is in this sense, the chameleon/symmetron
supported modified gravity models have limited scope for late time
cosmic acceleration.

\section{The hybrid symmetron model}

The preceding discussion clearly shows that modifications of
symmetron action (within the framework of conformal coupling at
least) would not help to satisfy local gravity constraints and
cosmological bounds suitable to late time evolution.

Late us note that the symmetron mass linearly depends upon the
density of the environment, the only parameter which remains to be
constrained by local gravity tests in this scenario is the mass
scale appearing in the coupling function. On the contrary, the
chameleon mass has a complicated non-linear dependence on density of
environment depending on the specific chameleon potential which
allows chameleon to satisfy the local physics constraints and also
play the roll of dark energy. As for the symmetron modification,
because the field is slowly  varying in time, we  have an
approximate equivalence between a modified symmetron action and the
standard symmetron. Thus within the framework of $\phi^4$ theory
with conformal coupling, the bound imposed by local physics on
symmetron mass cannot be improved. However, symmetron could still
play the role of cosmic phase transition facilitator, the role of
late time acceleration could be assigned to another field.

In what follows, we shall consider a model based upon the following
action,
\begin{align}
\label{hgene1}
S = &\int d^4 x
\sqrt{-g} \Bigl\{ \frac{R}{2\kappa^2} -\frac{1}{2}(\nabla\phi)^2-V(\phi) \nn
& + h(\phi) \mathcal{L}_\mathrm{dark\, energy} \Bigr\}
+\mathcal{S}_m
\left[ A^2(\phi)g_{\mu\nu},\Psi_m \right]\, .
\end{align}
where $h(\phi)$ is a coupling function which satisfies the conditions
\be
\label{Sy3} h(0) = 0\, ,\quad h(\phi_0)=1\, ,
\ee
for example, $h(\phi) =(\phi/\phi_0)^{m}$.

We may rewrite the
original symmetron potential in the following convenient form:
\be
\label{Sy1}
V(\phi) = \lambda \left( \phi^2 - \phi_0^2 \right)^2 \, .
\ee
Here $\phi_0^2=\mu^2/4\lambda$.
Notice that we added an effective constant ($\lambda \phi_0^4$) compared
to the potential previously studied.
As we will see, this constant will be always sub-dominant in cosmology
and hence do not contribute to the acceleration of Universe.

Let us note that in the high density regime such as the solar system
or the galaxy, the action (\ref{hgene1}) reduces to Einstein-Hilbert
one with a sub-dominant symmetron field. On the other hand, in
vacuum, the dark energy  contribution to the action dominates.


\subsection{$F(R)$ gravity switched by symmetron}

In this section, we propose a model where we use the symmetron to
generate the cosmic phase transition and the role of dark energy is
played by $F(R)$ gravity which is switched on by the symmetron
after the phase transition. This mechanism could alleviate some fundamental
problems like the radiative corrections to the mass of the scalaron (or more
generically, the mass of the quintessence field)
\cite{Upadhye:2012vh,Gannouji:2012iy} and the Frolov singularity \cite{Frolov:2008uf}.

The $F(R)$ gravity is a famous model to explain the accelerating expansion of the
present Universe.
A problem in the $F(R)$ gravity is the existence of the extra scalar mode which may give an
observable correction to the Newton law.
Such a correction could be observed by the solar system test, the experiments on the earth, etc.
In the present model, however, since the action reduces to that of the Einstein gravity
in the region in solar systems, in galaxy, or on the earth, the scalar mode does not appear and
could be consistent with any local test of the general gravity. That is, the symmetron screens the
extra scalar mode.

On the other hand, in the vacuum, there appears the $F(R)$ gravity,
which generates  accelerating expansion of Universe. A typical
example is \be \label{Sy7} F(R) = A_2 R^2\, . \ee Here $A_2$ is a
constant.

The $\Lambda$CDM model can be also realized by the $F(R)$ gravity without introducing the real
cold dark matter.
In the model with a cosmological constant and the matter with the EoS parameter $w$,
the scale factor behaves as $a=a_0\e^{g(t)}$ with
\be
\label{LCDM2}
g(t)=\frac{2}{3(1+w)}\ln \left(\mathcal{C} \sinh
\left(\frac{3(1+w)}{2l}\left(t - t_s \right)\right)\right)\, .
\ee
Here, $t_s$ is a constant of the integration and $\mathcal{C}$ is a constant.
The case $w=0$ corresponds to the $\Lambda$CDM model.
The behavior of the scale factor $a(t)$ can be realized by the $F(R)$ gravity model
by using the Gauss hypergeometric function \cite{Nojiri:2009kx}.

We should also note that the de Sitter space-time is an exact
solution of the wide class of the $F(R)$ gravity.

Furthermore, it has to clearly be emphasized that $F(R)$ theory is a kind of
scalar tensor theories. Indeed, it is well known that
the action of $F(R)$ theory
$
S_{F(R)} = \left(2\kappa^2\right)^{-1} \int d^4 x \sqrt{-g} F(R)
$
can be rewritten to
$
S_{\mathrm{ST}} = \int d^4 x \sqrt{-g} \left[ \left(2\kappa^2\right)^{-1}
\Phi R - W(\Phi) \right]
$ with $\Phi \equiv dF(R)/dR$ and
$W(\Phi) \equiv \left(2\kappa^2\right)^{-1}
\left[R(\Phi) \Phi - F(R(\Phi)) \right]$ a potential of $\Phi$.
Thus, the description in the formalisms of scalar tensor theories is
more generic than that in those of $F(R)$ theory.
In the following analysis,
since we would like to explicitly explore one of the theoretical
features of $F(R)$ gravity, we have directly introduced an $F(R)$ term.

In what follows, we investigate the profile of the scalar field
$\phi$ when the transition from the symmetric phase to the broken
phase is induced; we explain how the local gravity constraints
could be satisfied in the model under consideration.

We may consider the symmetron like potential (\ref{Sy1}).
Then the field equation is given by
\be
\label{profile1}
\Box \phi = 4\lambda \left( \phi^2 - \phi_0^2 \right) \phi + \alpha \rho \phi -h'(\phi)F(R)\, ,
\ee

where $\alpha=M^{-2}$ is the coupling to matter. We recover the previously
studied model when $F(R)=R$ and $m=2$.
In the following part, we do not consider this case.



We know that $F(R)\simeq H_0^2 M_\mathrm{Pl}^2$ which implies a negligible contribution
in the Klein Gordon equation. In fact when $\phi\simeq 0$, the $F(R)$ term is trivially negligible,
in the case where $\phi \simeq \phi_0$ we need to impose the condition $F(R)\ll \alpha \phi_0^2 \rho$.
We will define the range of viability for the parameters in order to satisfy this condition.
Assuming that the conditions are satisfied, the equation reduces locally to the standard form
without the $F(R)$-term.

We consider the static spherically symmetric solution in the flat space-time,
the equation has the following form:
\be
\label{profile2}
\frac{1}{r^2} \frac{d}{d r} \left( r^2 \frac{d \phi}{d r} \right)
= 4\lambda \left( \phi^2 - \phi_0^2 \right) \phi + \alpha \rho \phi\, .
\ee
We may assume
\begin{align}
\label{profile3}
\rho=\left\{ \begin{array}{ll}
\rho_0 & \mbox{when}\ 0\leq r<r_0 \\
0 & \mbox{when}\ r\geq r_0
\end{array} \right. \, .
\end{align}
We also assume $- 4\lambda \phi_0^2 + \alpha \rho_0 > 0$.

First, we consider the behavior of $\phi$ in the region $r\ll r_0$ and $\phi$ could be small.
Then we may linearize (\ref{profile2}) as
\be
\label{profile4}
\frac{1}{r^2} \frac{d}{d r} \left( r^2 \frac{d \phi}{d r} \right)
= \left( - 4\lambda \phi_0^2 + \alpha \rho_0 \right) \phi\, .
\ee
The solution is given by
\be
\label{ProfileA1}
\phi = \phi_\mathrm{in} \equiv
\frac{A \sinh \left( r \sqrt{- 4\lambda \phi_0^2 + \alpha \rho_0} \right)}{r} \, .
\ee
Here $A$ is a constant and we have assumed $\phi$ is finite at $r=0$.

Second, we consider the behavior of $\phi$ in the region $r\gg r_0$ and $\phi\sim \phi_0$.
By writing
\be
\label{ProfileA2}
\phi = \phi_0 + \delta \phi\, ,
\ee
we linearize the field equation (\ref{profile2}) with respect to $\delta \phi$ as follows,
\be
\label{ProfileA3}
\frac{1}{r^2} \frac{d}{d r} \left( r^2 \frac{d \delta \phi}{d r} \right)
= 8\lambda \phi_0^2 \delta \phi \, .
\ee
Then the solution is given by
\be
\label{ProfileA4}
\phi = \phi_\mathrm{out} \equiv \phi_0 - \frac{B \e^{- r \sqrt{8\lambda \phi_0^2}}}{r}\, .
\ee
In the region $r\ll r_0$, the solution (\ref{ProfileA1}) grows up very rapidly when $r$ increases.
Inversely if we fixed a value of $\phi$ for finite $r< r_0$, $\phi$ decreases very rapidly and goes to
vanish as $r$ decreases.

Different from the original symmetry model discussed in
Ref.~\cite{Hinterbichler:2010es,Hinterbichler:2011ca}, we may assume
the mass $\sqrt{8\lambda \phi_0^2}$ of the symmetron in the bulk
(low density regime) can be large. Then even in the region $r\gtrsim
r_0$, $\phi$ in the solution (\ref{ProfileA4}) goes to $\phi_0$ very
rapidly when $r$ increases. The above behaviors tells us that $\phi$
is almost constant except the small region $r\sim r_0$. Then in
order to estimate the constants $A$ and $B$, we match the solutions
$\phi_\mathrm{in}$ and $\phi_\mathrm{out}$ by imposing the following
boundary conditions \be \label{ProfileA5} \phi_\mathrm{in}(r_0) =
\phi_\mathrm{out}(r_0) \, , \quad \phi_\mathrm{in}'(r_0) =
\phi_\mathrm{out}'(r_0) \, . \ee We find
\begin{widetext}
\begin{align}
\label{ProfileA6}
A =& \frac{\phi_0 \left( 1 + r_0 \sqrt{8\lambda \phi_0^2} \right)}
{\sqrt{8\lambda \phi_0^2} \sinh \left( r \sqrt{- 4\lambda \phi_0^2 + \alpha \rho_0} \right)
+ \sqrt{- 4\lambda \phi_0^2 + \alpha \rho_0}
\cosh \left( r \sqrt{- 4\lambda \phi_0^2 + \alpha \rho_0} \right)}\, , \nn
B =& \frac{\phi_0 r_0 \e^{r \sqrt{8\lambda \phi_0^2}}
\left( - \frac{1}{r_0} \sinh \left( r \sqrt{- 4\lambda \phi_0^2 + \alpha \rho_0} \right)
+ \sqrt{- 4\lambda \phi_0^2 + \alpha \rho_0}
\cosh \left( r \sqrt{- 4\lambda \phi_0^2 + \alpha \rho_0} \right)\right)}
{\sqrt{8\lambda \phi_0^2} \sinh \left( r \sqrt{- 4\lambda \phi_0^2 + \alpha \rho_0} \right)
+ \sqrt{- 4\lambda \phi_0^2 + \alpha \rho_0}
\cosh \left( r \sqrt{- 4\lambda \phi_0^2 + \alpha \rho_0} \right)}\, .
\end{align}
\end{widetext}
We may assume $r_0$ could be a radius of planet, star, or galaxy.
Then $r_0$ should be much larger than the Compton lengths $1/\sqrt{8\lambda \phi_0^2}$
and $1/ \sqrt{- 4\lambda \phi_0^2 + \alpha \rho_0}$.
Then since $\sinh \left( r \sqrt{- 4\lambda \phi_0^2 + \alpha \rho_0} \right)
\sim \cosh \left( r \sqrt{- 4\lambda \phi_0^2 + \alpha \rho_0} \right)
\sim \e^{ r \sqrt{- 4\lambda \phi_0^2 + \alpha \rho_0}}/2$, we find
\begin{align}
\label{ProfileA7}
A &\sim \frac{2\phi_0 r_0 \e^{- r_0\sqrt{- 4\lambda \phi_0^2 + \alpha \rho_0}} }
{1 + \sqrt {\frac{\alpha \rho_0}{8\lambda \phi_0^2} - \frac{1}{2}}}\, ,\\
B &\sim \frac{\phi_0 r_0 \e^{ r_0\sqrt{8\lambda \phi_0^2 }}
\sqrt {\frac{\alpha \rho_0}{8\lambda \phi_0^2} - \frac{1}{2}}}
{1 + \sqrt {\frac{\alpha \rho_0}{8\lambda \phi_0^2} - \frac{1}{2}}}\, .
\end{align}
We now consider the strength of the force $F_\phi$ generated by the scalar field $\phi$.
Since the strength of the coupling is given by $\alpha\phi$, we find
\be
\label{ProfileA8}
F_\phi (r) = m \alpha\phi \frac{d \phi}{d r}\, .
\ee
Here $m$ is the mass of the particle receiving the force.
In the region $r\gg r_0$, by using (\ref{ProfileA4}),
we find
\begin{align}
\label{ProfileA9}
\frac{F_\phi (r)}{m} =& \alpha B \left(
\phi_0 - \frac{B \e^{- r \sqrt{8\lambda \phi_0^2}}}{r}\right)
\left(\frac{\sqrt{8\lambda \phi_0^2}}{r} + \frac{1}{r^2}
\right) \nn
& \times \e^{- r \sqrt{8\lambda \phi_0^2}} \nn
\sim & \frac{\alpha B \phi_0^2 \sqrt{8\lambda}}{r}
\e^{- r \sqrt{8\lambda \phi_0^2}} \, .
\end{align}
On the other hand, the Newtonian force of gravity $F_g$ is given by
$F_g (r) = GM m/r^2$, where $M\equiv 4 \pi \rho_0 r_0^3/3$. We
notice that in the region specified by $r\gg r_0$, the force
$F_\phi$ generated by the scalar field $\phi$ can be neglected
compared to the Newtonian force.

\subsection{Constraints on model parameters \label{S10}}

We now consider the constraints for the parameters.
A constraint comes from the condition that the symmetry is restored
when $r<r_0$ but broken in the vacuum:
\be
\label{CC1}
\alpha \rho_0 > 2\lambda \phi_0^2 > \alpha \rho_\infty \sim \alpha H_0^2 M_\mathrm{Pl}^2\, .
\ee
Here $\rho_\infty$ is the energy density of the vacuum, $H_0$ is the Hubble constant in the present Universe,
and $M_\mathrm{Pl}$ is the Planck mass.
Another constraint may come from the condition that the Compton length of the scalar field
even in the vacuum should be much smaller than the size of galaxies $r_\mathrm{G}$:
\be
\label{CC2}
\frac{1}{\sqrt{8\lambda \phi_0^2}} \ll r_\mathrm{G}\, .
\ee
We also require that $h(\phi)F(R)$ term does not affect the phase transition,
\be
\label{CC3}
V(0) = \lambda \phi_0^4 \gg F(R_\infty) \sim H_0^2 M_\mathrm{Pl}^2\, .
\ee
In case of the galaxy $\rho_0 \sim 10^5\, H_0^2 M_\mathrm{Pl}^2$ and
$r_\mathrm{G} \sim \left( 10^{-7}-10^{-5} \right) \, H_0^{-1}$.

Since $H_0 \sim 10^{-61} M_\mathrm{Pl}$ and $G=8\pi M_\mathrm{Pl}^2$,
by introducing the new variables
\be
\label{CC4}
x= \phi_0^2\, ,\quad y = \lambda \phi_0^2\, ,\quad z = \frac{1}{\alpha}\, ,
\ee
and putting $r_0 = r_\mathrm{G}$,
the above constraints (\ref{CC1}), (\ref{CC2}), and (\ref{CC3}) can be
rewritten as
\begin{align}
\label{CC5}
yz &\ll 10^{-117}M_\mathrm{Pl}^4\, , \quad
yz \gg 10^{-122}M_\mathrm{Pl}^4\, ,\\
y &\gg 10^{-108}M_\mathrm{Pl}^2\, ,\quad
xy \gg 10^{-122}M_\mathrm{Pl}^4\, .
\end{align}
By using, $z=10^n M_\mathrm{Pl}^2$, we obtain
\begin{align}
\label{CCC1}
y &\ll 10^{-117 - n} M_\mathrm{Pl}^2\, ,\quad
y \gg 10^{-122 - n} M_\mathrm{Pl}^2\, ,\\
y &\gg 10^{-108} M_\mathrm{Pl}^2\, ,\quad
xy \gg 10^{- 122} M_\mathrm{Pl}^4\, .
\end{align}
We find  $xy \gg 10^{- 122} M_\mathrm{Pl}^4$, and for $-14<n<-9$
\begin{align}
\label{CCC21}
10^{-117-n} M_\mathrm{Pl}^2 \gg y \gg 10^{-108} M_\mathrm{Pl}^2\, ,~ x \gg 10^{-5+n} M_\mathrm{Pl}^2 \, ,
\end{align}
and for $n<-14$
\begin{align}
\label{CCC22}
10^{-117-n} M_\mathrm{Pl}^2 \gg y \gg 10^{-122-n} M_\mathrm{Pl}^2\, , ~ x \gg 10^{-5+n} M_\mathrm{Pl}^2 \, .
\end{align}
 We therefore confirm  that we have always a range of viability of the parameters,
where all the constraints are satisfied.

\subsection{Phase transition in the early Universe \label{S11}}

As mentioned earlier, the phase transition should have occurred in
rather early Universe, not in the late Universe. We now investigate
how the transition could have occurred and if there could be any
problem or not. The potential (\ref{Sy1}) tells that the critical
density $\rho_\mathrm{cr}$, where the phase transition occurs, is
given by \be \label{CCC3} \rho_\mathrm{cr} = \frac{4\lambda
\phi_0^2}{\alpha} = 4 yz \, . \ee When the phase transition could
have occurred, the Hubble rate $H_\mathrm{cr}$ is given by \be
\label{CCC4} H_\mathrm{cr}^2 = \frac{\kappa^2}{3} \rho_\mathrm{cr}
\sim \frac{yz}{M_\mathrm{Pl}^2}\, . \ee Here we have used the
notation defined in (\ref{CC4}). As we are interested in the
behavior when $\phi\sim \phi_0$, by using (\ref{ProfileA2}), we
linearize the scalar field equation as \be \label{CCC5} \frac{d^2
\delta \phi}{dt^2} + 3 H_\mathrm{cr} \frac{d \delta \phi}{dt} + 8
\lambda \phi_0^2 \delta \phi =0\, . \ee Equation~(\ref{CCC5}) has a
form analogous to the equation of motion of the harmonic oscillator
with drag (air resistance): \be \label{CCC6} \frac{d^2 x}{dt^2} +
2\gamma \frac{d x}{dt} + \omega^2 x =0\, . \ee Here $\gamma$ and
$\omega$ are positive constants and $x$ expresses the position of
the harmonic oscillator. As well-known, when $\gamma^2 \geq
\omega^2$, the amplitude $|x|$ decreases without oscillation and
when $\gamma^2 < \omega^2$, the amplitude $|x|$ decreases with
oscillation by the angular velocity $\sqrt{\omega^2 - \gamma^2}$.
Hence the time scale relevant to the decrease of amplitude is given
by $T_\mathrm{dec} = 1/\gamma$. Then Eq.~(\ref{CCC5}) tells us that
$\delta \phi$  would vanish without oscillation if $9
H_\mathrm{cr}^2 = \frac{9yz}{M_\mathrm{Pl}^2}> 8 \lambda \phi_0^2=8
y$, that is, $\frac{z}{M_\mathrm{Pl}^2} \gtrsim 1$. On the other
hand, $\delta \phi$  vanishes with oscillation if
$\frac{z}{M_\mathrm{Pl}^2} \lesssim 1$. By writing $z=10^n
M_\mathrm{Pl}^2$ as before, and since $n<-9$, we find $\delta \phi$
 vanishes with oscillation. The time scale of decreasing the
amplitude is given by \be \label{CCC7} T_\mathrm{dec} = 1/H \sim
M_\mathrm{Pl} \sqrt{yz}\, . \ee Then Eq.~(\ref{CCC21},\ref{CCC22})
tells when $-14<n<-9$
\begin{align}
\label{CCC8}
10^{58} M_\mathrm{Pl}^{-1} \ll T_\mathrm{dec} \ll  10^{54 -n/2} M_\mathrm{Pl}^{-1}\, ,
\end{align}
and for $n\leq -14$
\begin{align}
10^{58} M_\mathrm{Pl}^{-1} \ll T_\mathrm{dec} \ll 10^{61} M_\mathrm{Pl}^{-1} \, .
\end{align}
Since $M_\mathrm{Pl}^{-1} \sim 10^{-44}\, \mathrm{sec} \sim 10^{- 51}\, \mathrm{years}$,
we have for $-14<n<-9$
\begin{align}
\label{CCC9}
10^{7}\, \mathrm{years} \ll T_\mathrm{dec} \ll 10^{3 - n/2}\, \mathrm{years}\, ,
\end{align}
and for $n \leq -14$
\begin{align}
\label{CCC91}
10^{7}\, \mathrm{years} \ll T_\mathrm{dec} \ll 10^{10}\, \mathrm{years}\, .
\end{align}
We therefore conclude that it could take 10 million as minimum or 10
billion years as maximum for decreasing the oscillations. The period
$T_\mathrm{P}$ of the oscillation is given by $2\pi /\sqrt{8 \lambda
\phi_0^2 } \sim 1/\sqrt{y}$. Therefore Eqs.~(\ref{CCC9},\ref{CCC91})
 gives
\begin{align}
\label{CCC10}
& 10^{7 + \frac{n}{2}}\, \mathrm{years}  \ll T_\mathrm{P} \ll 10^{3}\, \mathrm{years}\, ,\\
& 10^{7 + \frac{n}{2}}\, \mathrm{years} \ll T_\mathrm{P} \ll 10^{10 + \frac{n}{2}}\, \mathrm{years}\, .
\end{align}

 for $-14<n<-9$ and $n\leq -14$ respectively. For instance, if we
choose $n=-14$, we have $ 1\, \mathrm{year} \ll T_\mathrm{P} \ll
10^3\, \mathrm{years}$. Therefore the oscillations are very slow and
do not give rise to particle production.

Although the phase transition did not occur during the accelerating
expansion, that is, when the redshift is $z\sim 1$, we may estimate
when it could have occurred. Since we assume $\rho_0 \sim 10^5\,
H_0^2 M_\mathrm{Pl}^2 > \rho_\mathrm{cr} > \rho_\infty$ and $\rho
\propto a^{-3}$, we may assume the phase transition could have
occurred when the redshift $z=\frac{a_0}{a} - 1 \sim 10$ ($a_0$ is
the present value of the scale factor $a$), which corresponds to a
few million years after the Big-Bang; the redshift of the cosmic
microwave background radiation corresponds to $z\sim 1000$.

In the early universe, the $Z_2$ symmetry could be restored. Then
Eq.~(\ref{Sy1}) tells us that $V(\phi)$ gives a contribution to the
energy in the vacuum by $\lambda \phi_0^4$. The contribution is,
however, not so large. Equation~(\ref{CC4}) tells $\lambda \phi_0^4
= xy$. Using the previous results, we find $\lambda \phi_0^4 = x y >
10^{-122}\, \left( M_\mathrm{Pl} \right)^4$. On the other hand, for
an example, the energy density of the matter in the present universe
is $10^{-124}\, \left( M_\mathrm{Pl} \right)^4$. When $z>10$, since
the matter density could be $10^{-124}\times \left(1+z\right)^3 \,
\left( M_\mathrm{Pl} \right)^4 \sim 10^{-121}\, \left( M_\mathrm{Pl}
\right)^4$ Then the magnitude of $V(0) = \lambda \phi_0^4$ is
comparable with the energy density of matter but could not be
dominant. Therefore the contribution from the constant $\lambda
\phi_0^4 = x y$ can always be neglected.

\section{Summary and Discussion \label{S16}}

In this paper, we examined the modified symmetron models with an aim
to reconcile the latter with late time cosmic acceleration. We have
investigated a symmetron type framework coupled to curvature scalar
and to matter. In this case the effective symmetron potential
drastically differs from the one in original scenario. We examined
the consistency of the model with local physics and found
constraints on the symmetron mass similar to the one in the standard
 model. The latter is related to the fact that local gravity tests
are insensitive to the details of the effective symmetron potential
and the new mass scale appearing in the over all coupling in the
leading approximation ($\phi/M\ll 1\, ,\ \phi/M_\mathrm{Pl}\ll 1$)
in the model under consideration is same as the one that defines the
symmetron mass. We therefore conclude that there is a ``no go''
theorem for dark energy symmetron as long as we confine to conformal
coupling, situation might(might not) change in the disformal case.

Being inspired by the beauty of cosmic symmetry breaking, we have
made a proposal in which symmetron felicitates the phase transition
in low matter density regime ; the symmetry is restored in high
density ensuring the compliance of the model with local gravity
constraints. The role of symmetron field ceases after phase
transition thanks to its coupling to an $F(R)$ gravity action or
quintessence field.


In case of $F(R)$ symmetron or equivalently a quintessence field,
the model reduces to the standard Einstein gravity in the high
density region whereas $F(R)$ gravity (or quintessence) is switched
on in the low density region; the transition is generated by the
symmetron field. Since in our proposal, symmetron field is
responsible for phase transition only, its mass could be large even
in the vacuum and therefore the local gravity constraint can easily
be satisfied, which is different from the original symmetron model
\cite{Hinterbichler:2010es,Hinterbichler:2011ca}. We have also
investigated other constraints and we have shown that there exists a
parameter region which satisfies all constraints. In the model under
consideration, the transition could have occurred around $z\sim 10$
-- the dark age. After the transition, symmetron field oscillates
around the true minimum but the time scale of decreasing of
amplitude varies between 10 million to 10 billion years which is
much smaller compared to the Hubble scale. On the other hand, the
period of the oscillations varies from one to one thousand years,
which could be large enough from thermodynamical point of view. Thus
the oscillation would not conflict with the observed cosmology.

Last but not least, it might be interesting to enlarge the symmetron
framework to disformal type of set up to check whether the disformal
symmetron can reconcile with dark energy.

\section*{Acknowledgments.}

We are grateful to J.~Matsumoto for the discussion.
K.B. would like to sincerely appreciate National Center for Theoretical
Sciences and National Tsing Hua University very much for the very
kind and warm hospitality, where this work has been finalized.
S.N. is supported by Global COE Program of Nagoya University (G07)
provided by the Ministry of Education, Culture, Sports, Science \&
Technology and by the JSPS Grant-in-Aid for Scientific Research (S) \# 22224003
and (C) \# 23540296.
M.S. is supported by the JSPS Invitation Fellowship for Research
in Japan (Long-Term) \# L12422 and also
by the Department of Science and Technology, India
under project No. SR/S2/HEP-002/2008. R.G. is supported by the Grant-in-Aid for Scientific
Research Fund of the JSPS No. 10329

\end{document}